# Mass spectrometric investigations into 3D printed parts to assess radiopurity as ultralow background materials for rare event physics detectors

*French AD, Anguiano SA, Bliss M, Christ J, di Vacri ML, Erikson R, Harouaka K, Hoppe EW, Grate JW, Arnquist IJ**

*corresponding author

**Abstract**

The mass concentrations of $^{232}$Th and $^{238}$U in several 3D printing filaments and printed polymer parts are reported as measures of their radiopurity. To minimize background signals in rare event physics detectors, radiopure polymers are necessary as dielectric and structural materials in their construction. New data are reported for polyvinylidene fluoride (PVDF), polyphenylene sulfide (PPS), and two forms of polyetherimide (PEI, branded ULTEM 1010 and 9085). Data for starting filaments and both simple and complex printed parts are reported. PVDF filaments and simple printed beads, were found to have values of approximately 30 and 50 pg g$^{-1}$ for $^{232}$Th and $^{238}$U, respectively, while a more complex spring clip part had slightly elevated $^{232}$Th levels of 65 pg g$^{-1}$, with $^{238}$U remaining at 50 pg g$^{-1}$. PPS filament was found to have concentrations of 270 and 710 pg g$^{-1}$ for $^{232}$Th and $^{238}$U, respectively, and were not chosen to be printed as those levels were already higher than other material options. ULTEM 1010 filaments and printed complex spring clip parts were found to have concentrations of around 5 and 7 pg g$^{-1}$ for $^{232}$Th and $^{238}$U, respectively, illustrating no significant contamination from the printing process. ULTEM 9085 filaments were found to have concentrations of around 9 and 5 pg g$^{-1}$ for $^{232}$Th and $^{238}$U, respectively, while the printed complex spring clip part was found to have slightly elevated concentrations of 25 and 7 pg g$^{-1}$ for $^{232}$Th and $^{238}$U, respectively. These results were all obtained using a novel dry ashing method in crucibles constructed of ultralow background electroformed copper or, when applicable, microwave assisted wet ashing digestion. Samples and process blanks were spiked with $^{229}$Th and $^{233}$U as internal standards prior to dry/wet ashing and determinations were made by inductively coupled plasma mass spectrometry (ICP-MS). In order to maintain high radiopurity levels, pre-cleaning the filaments before printing and post-cleaning the parts is recommended, although the printing process itself has shown to contribute very minute amounts of radiocontaminants.

# 1. Introduction

Additive manufacturing (3D printing) provides a way to produce a variety of complex devices at a reduced cost while minimizing waste and production time [1–3]. The utility of 3D printing has been exploited for a variety of applications, including architecture, engineering, medical/dental, and construction [1-2] . Over the past five years 3D printing has rapidly become a faster and cheaper option for testing prototypes in many industries. For example, 3D printing has been utilized by analytical laboratories to test novel part designs without the cost and lead times required for custom mold or nozzle manufacture.

Typical methods of 3D printing include fused deposition modeling (FDM), stereolithography (SLA), selective laser sintering (SLS), and selective laser melting (SLM), among others [5]. The most common and cost-effective method is FDM which utilizes a thermoplastic polymer filament that is fed into a heated nozzle and extruded in layers, solidifying at room temperature to create the desired part. There are a wide variety of thermoplastics that can be used in FDM 3D printing such as polycarbonate (PC), polyethylene terephthalate (PET), and high-density polyethylene (HDPE), just to name a few.

The work presented herein is a preliminary investigation into the utility of 3D printed parts as ultralow background materials for rare event physics detectors. Such detectors are used in the search for dark matter, solar neutrinos and neutrinoless double beta decay, where the background signals of various radioactive elements must be reduced. Ideally, parts used to construct these detectors would have little-to-no background radioactivity that would generate signals that interfere with the rare event signal of interest. For 3D printing to be used for preparation of ultralow background (ULB) detector parts, the part needs to be extremely radiopure in the primordial radionuclides $^{238}$U and $^{232}$Th (*e.g., ca.* microBq kg$^{-1}$, or pg g$^{-1}$ levels). The printers are easy to access, and offer more control over the manufacturing process, particularly for very complex parts that may challenge traditional manufacturing methods. 3D printing has the potential to reduce the need for dedicated equipment and clean machining which can be a cumbersome and expensive process. 3D printing could also reduce the mass of some parts by including void volumes thus lowering overall background. The printers are easy to implement in underground cleanrooms providing the option to prepare parts on-site as cleanly as possible when needed. A first step into understanding applications of 3D printed polymers is sourcing and assaying starting filaments and printed parts to understand if viable materials can be produced to meet the stringent radiopurity requirements of next-generation ULB physics experiments. Previous investigations in our group showed very promising results for the starting FDM filaments for ULTEM

1010, at microBq kg$^{-1}$ levels [6]. This study goes further to understand if *printed* parts can be attained at similarly clean levels.

Here we focus on four different high-performance polymers used as filaments in a FDM 3D printer to investigate the radiopurity of each polymer before and after printing and the potential utility of 3D printing for ULB physics detectors. Assay results for one polymer (polyvinylidene fluoride, PVDF) from 3D printing were compared to polymers assayed previously [4] to provide a general comparison to polymers at different stages of sourcing (stock powder, stock pellets, stock filaments, and bulk parts). As far as the authors are aware, this is the first study to source and investigate the radiopurity of 3D printed parts for ULB applications. While some polymers were found to contain relatively high concentrations of Th or U in either the starting material or printed parts, our data demonstrates that some polymer types can be printed cleanly, without substantially increasing radiopurity at levels sufficiently radiopure for ULB applications.

## 2. Methodology

*2.1 Materials*

Four different starting materials were investigated: polyvinylidene fluoride (PVDF), two polyetherimide materials (PEI, name brand ULTEM 1010 and 9085), and polyphenylene sulfide (PPS). All filaments were of 1.75 mm diameter and purchased through 3DXTECH (Grand Rapids, MI, USA). These polymers, specifically PVDF and ULTEM, were selected based on their superior mechanical properties and have been shown to be sourced from radiopure stock [4-5].

Sample preparation and analyses were performed in a Class 10,000 cleanroom at Pacific Northwest National Laboratory (PNNL). A laminar flow hood providing a Class 10 environment was used for sample preparation. Sampling implements (*e.g.*, forceps) and PFA vials (Savillex, Eden Prairie, MN) that contacted any part of the sample during handling were first leached in 6 M HNO$_3$ and then validated for their cleanliness. Validations involved leaching the labware in a 5% (v/v) HNO$_3$ solution before analyzing the leachate via ICP-MS. Implements passed validation if the leachate was verified to be at reagent blank levels.

*2.2 3-D printing procedure*

Samples were printed on a custom printer platform modified to support high temperature fabrication. The custom printer was modified from a MakerBot Replicator 2X (MakerBot Industries,

Brooklyn, NY) and utilizes Repetier-Host as the controlling software and Slic3r as the slicing software. These machines typically store the feed spool in an open chamber. For this study, the feed spool was in a closed plastic box and fed through PTFE tubing to just before the printing head. This protected the filament from dust and dirt in the room. Some polymers require a "raft" material to adhere the molten polymer to the printing deck to initiate part formation. Initially, small beads were printed (Figure 1a) to test the polymer cleanliness for a simple design, then a spring clip was printed (Figure 1b) as a test piece representing a realistic and complex part. The materials were printed using the parameters listed in Table 1.

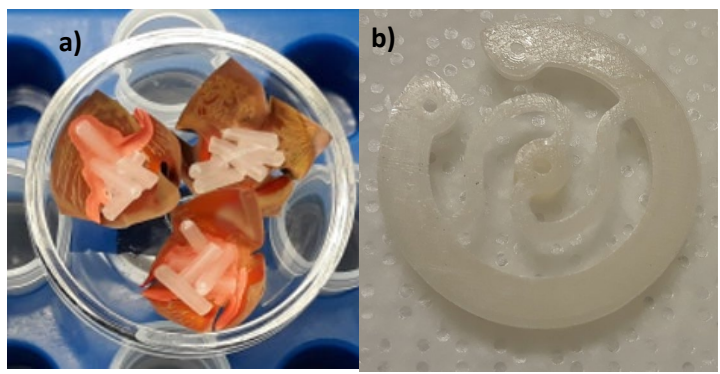

**Figure 1.** 3D printed PVDF beads in an electroformed Cu crucible prior to dry ashing (a) and PVDF spring clip prior to subsampling and dry ashing (b).

**Table 1.** Printing parameters used for the various materials tested.

| Material | Nozzle Temp. (°C) | Bed Temp. (°C) | Nozzle | Raft |
|---|---|---|---|---|
| PVDF | 260 | 110 | Brass | No |
| ULTEM 1010 | 390 | 120 | Stainless | Yes |
| ULTEM 9085 | 370 | 115 | Stainless | Yes |
| PPS | 345 | 110 | Brass | No |

To ensure that only clean feedstock was incorporated into the printed samples, the system was purged by feeding clean feedstock into the extruder until the sample appeared visually clean. An additional 200mm of filament was then purged through the system before the print was started. Two sets of PVDF samples were printed, using 400μm and 200μm brass nozzles. The ULTEM samples were printed using a 400μm steel nozzle due to the high print temperature needed and were printed onto a constructed removable raft of the same material. The 400μm nozzle samples were printed at 150μm layer heights, the 200μm nozzle samples were printed at 100μm layer heights. No PPS samples were printed, but the processing temperatures are included for comparison. Samples were printed at average speeds of 15-20 mm/s.

*2.3 Sample Preparation*

The starting filaments and final printed products were cleaned by sonicating in a 2% (v/v) micro90 (Cole-Parmer, Vernon Hills, IL) solution for 10 minutes, followed by a brief (*ca.* 1-2 min) sonication in 6M $HNO_3$ (Optima Grade; Fisher Scientific, Pittsburgh, PA). Sonication in $HNO_3$ was kept to a minimum for ULTEM and PPS samples to minimize surface corrosion. All parts received a final rinse in DI water and were digested using Cu crucible dry-ashing or microwave digestion described below. Subsamples were on the order of 30-150 mg, and most cases each part or filament was measured in triplicate.

Sample dry ashing was performed using Cu crucibles following the procedure outlined in Arnquist et al., [7]. In short, Cu crucibles were utilized as the sample vessel for dry ashing to maintain cleanliness and good recoveries of analytes. For quantitation, a spike of $^{229}$Th and $^{233}$U (Oak Ridge National Laboratory, Oak Ridge, TN) was added to each process blank (empty copper crucible) and sample. Process blanks and each unique polymer (filament, printed bead, or printed part) were assayed in triplicate. Boats were placed in the quartz tube furnace and ashed using a controllable temperature heating program. Samples were ashed using a slow-ramping heating program (14 h, overnight) with a maximum temperature of 800 °C in the presence of air at a flow rate of 4 L/min. After ashing, the (now) oxidized copper boats and remaining residual metals were retrieved and moved to their respective validated PFA vial. The copper boats were then digested in 8 M $HNO_3$ to a final copper concentration of 0.175 g/mL solution. Pre-concentration and matrix removal of the dissolved copper was accomplished from the adapted procedure [8] and analyzed via ICP-MS.

Microwave digestion was performed using a Mars 6™ microwave digestion system with iPrep™ vessels (CEM Corp., Matthews, NC) on some of the ULTEM samples, as these samples could also be digested using wet ashing techniques (PVDF is impervious to acid attack and requires dry ashing). Samples were loaded into cleaned and validated vessels, along with a known amount of $^{229}$Th and $^{233}$U radiotracer solution, and 5 mL of $HNO_3$. Process blanks were prepared in triplicate using the same procedure. The heating program used a 30 min heat ramp to 250 °C and a hold at this temperature for

30 min before cooling to room temperature. Fully digested sample solutions were retrieved, transferred to cleaned/validated PFA vials, and reconstituted into 1.8 mL of 2% $HNO_3$ before analysis via ICP-MS.

*2.5 Instrumentation and data analysis*

An Agilent 8800s series ICP-MS (Agilent Technologies, Santa Clara, CA), equipped with an integrated autosampler, standard quartz double-pass spray chamber and a microflow nebulizer was used for determination of $^{232}$Th and $^{238}$U. Tuning of instrument operating parameters, including plasma, ion optics, and mass analyzer settings, was performed daily using a 0.1 ng/g Tl standard to maximize sensitivity at the high *m/z* range to optimize signal-to-noise for Th and U.

Quantification was performed using isotope dilution methods where all process blanks and samples were spiked with a known amount of $^{229}$Th and $^{233}$U radiotracer before dry/wet ashing. A *ca.* 100 fg g$^{-1}$ $^{229}$Th and $^{233}$U standard was used to monitor signal intensity throughout the analysis. Integration times were typically 5-15 s for each isotope of interest (i.e., $^{229}$Th, $^{232}$Th, $^{233}$U, and $^{238}$U) with three signal acquisitions performed per *m/z* for each sample. Relative standard deviations for $^{232}$Th and $^{238}$U were *ca.* 5%.

**3. Results and Discussion**

*3.1 Investigations into PVDF*

PVDF is a strong fluoropolymer that is typically used in the semiconductor manufacture industry and has been previously investigated for radiopurity in several forms: powder, pellet and solid [7,9]. Previous work showed that radiopurity of the different forms worsened with additional processing steps. That is, the radiocontaminants in the powder < pellets << bulk formed part. The powder, which goes through the fewest processing steps, was found to contain concentrations of < 2 pg g$^{-1}$ Th and U, while the pellets, which are created from the powder, contained *ca.* 3 and 11 pg g$^{-1}$ Th and U, respectively. Finally, the formed bulk PVDF parts were much dirtier (*ca.* 30 and 390 pg g$^{-1}$ Th and U, respectively), which is thought to be due to additional handling during material processing at the manufacturer. Given the encouragingly clean radiopurity values for the stock PVDF materials, the PVDF filament (which is sourced from the same clean supply chain as that of the powder, pellet, and bulk part) was investigated in this study. The thought was that if the PVDF filaments were sufficiently radiopure, then perhaps clean additive manufacturing processes could be developed (and verified through assay) to maintain cleanliness in the printed part. Tabulated assay results for the starting filaments and printed parts are shown in Table 2, and a visual representation is shown in Figure 2.

Table 2. Average Th and U content (both in pg g$^{-1}$ and μBq kg$^{-1}$) of PVDF filaments and printed parts. Each polymer was measured in triplicate using the copper crucible dry ashing method.

| Sample | Description | $^{232}$Th pg g$^{-1}$ | $^{232}$Th μBq kg$^{-1}$ | $^{238}$U pg g$^{-1}$ | $^{238}$U μBq kg$^{-1}$ |
|---|---|---|---|---|---|
| PVDF | Cleaned Filament | 30.5 ± 1.2 | 125 ± 5 | 49 ± 5 | 610 ± 60 |
| | | | | | |
| PVDF | Cleaned Bead from Cleaned Filament | 31 ± 14 | 130 ± 60 | 50 ± 20 | 600 ± 300 |
| PVDF | Uncleaned Bead from Cleaned Filament | 31 ± 3 | 126 ± 12 | 43 ± 17 | 500 ± 200 |
| | | | | | |
| PVDF | Cleaned Spring Clip from Cleaned Filament | 65 ± 5 | 270 ± 20 | 52 ± 8 | 650 ± 100 |
| PVDF | Cleaned Spring Clip from Uncleaned Filament | 300 ± 400 | 1300 ± 1700 | 90 ± 40 | 1200 ± 500 |

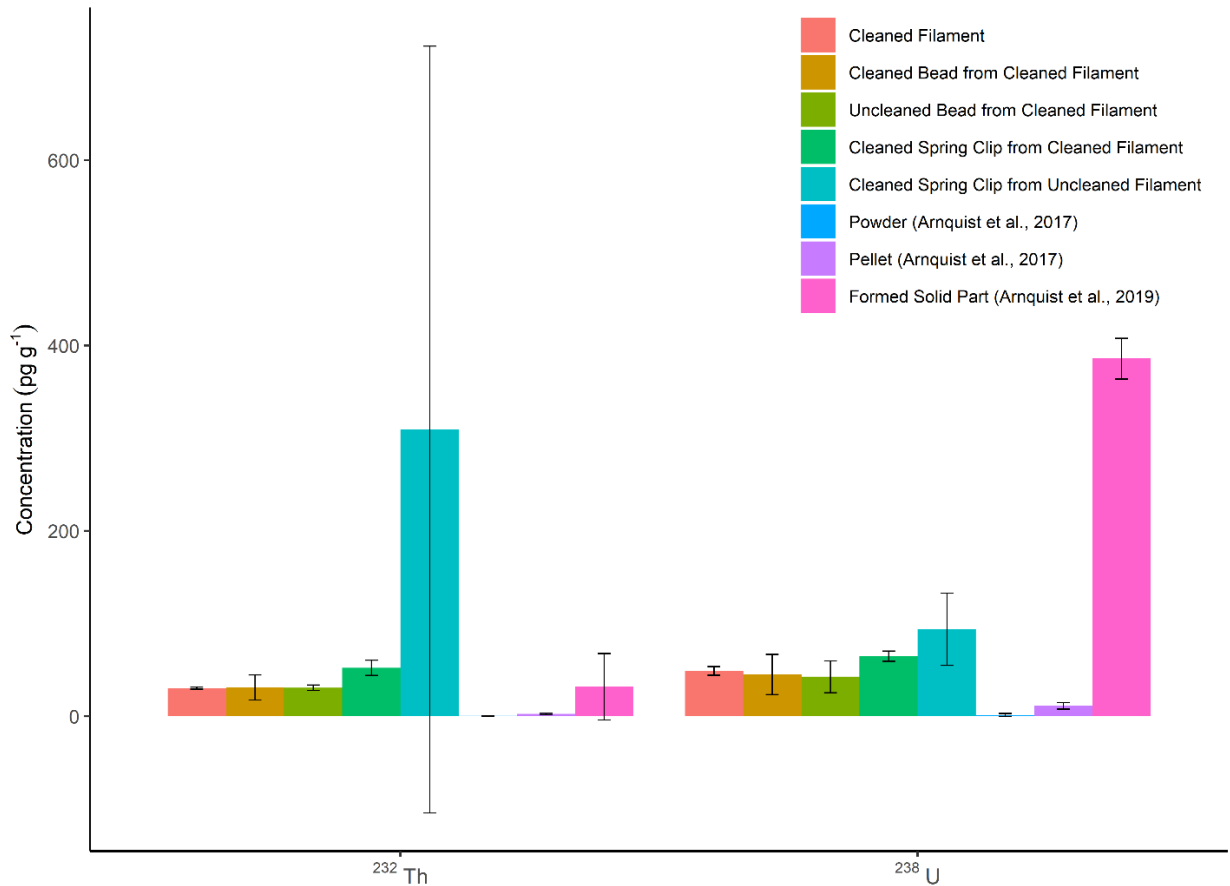

Figure 2. Average $^{232}$Th and $^{238}$U concentrations (pg g$^{-1}$) in PVDF filament, 3-D printed bead, and 3-D printed Spring Clip, as well as results from previous work [7,9] for comparison. Error bars represent the standard deviation of three replicates

Following the filament assays a "bead" of PVDF was printed to investigate how a simple 3D printing procedure would affect the polymer radiopurity. Moreover, to assess if any radiocontaminants were introduced at/near the surface of the printed bead, two subsets of beads were assayed, one set that was cleaned as described in the Experimental, and one set that was assayed "as received" from the 3D printer.  Both subsets (the cleaned and uncleaned) were printed directly into a cleanroom poly bag to minimize any extraneous handling at the non-cleanroom print shop before being taken to the analytical chemistry cleanroom (Class 10,000) for assay.  Results show there was no difference between the washed and unwashed beads from the printer, illustrating that the printing process contributes negligible amounts of radiocontaminants when printing a small part with a simple design.

To determine if more intricate printing procedures introduce additional contamination the PVDF filament was then used to print a more complex part, a proportional counter spring clip (Figure 1b). While there was no significant difference in U levels, the spring clip had roughly twice as much Th after printing than the simple beads and stock filament.  When the spring clip was printed from an uncleaned filament the average Th contamination increased by 10x compared to the simple bead and the average U concentration doubled. It is most likely that the external contamination on the filament is incorporated into the part during the printing process and additional cleaning will not remove this contamination.  The high variability across replicates (standard deviations) also reflects the variable nature of external contamination on the uncleaned filament.  Thus, it is recommended that all filaments be cleaned prior to printing complex parts as any residual contaminants may get incorporated into the "bulk" during the printing process.

In summary of the PVDF results, the PVDF filament obtained for this study was cleaned initially to remove any external contamination and analyzed prior to printing and found to contain an average of 30.5 and 49 pg g$^{-1}$ Th and U, respectively.  While the radiopurity in the PVDF filament is not as good as previously determined for the stock PVDF powder and pellets—it is likely there is an additional step to manufacture the filaments for 3D printing that introduces an additional source of contamination—the levels in the 3D printed beads and complex spring clip parts printed from cleaned filaments are very encouraging compared to radiopurity levels seen in bulk PVDF parts previously, and are very close to the radiopurity of the precursor filaments themselves. The 3D printed PVDF spring clip was more radiopure than the cleaned bulk formed PVDF parts [9] , showing that 3D printed bulk parts are attainable at cleaner levels than those formed using conventional methods.

*3.2 Investigations into PPS and ULTEM*

Following the initial printing investigations with the PVDF polymer, three other polymer printing investigations were conducted on PPS, ULTEM 1010, and ULTEM 9085. While the PPS filament contained the greatest concentrations of Th (270 ± 60 pg g$^{-1}$) and U (710 ± 20 pg g$^{-1}$) in the starting filaments and was not utilized in further testing, the two types of ULTEM showed encouraging results.

ULTEM is known for having a high mechanical strength and is generally stronger than fluorinated or chlorinated thermoplastics and has showed promise for use as an electronic interposer in ultralow background detectors, among other applications. ULTEM 1010 is pure PEI while ULTEM 9085 contains a polycarbonate (PC) copolymer which is added for improved flow. Previous work into ULTEM 1010 filament investigations showed promising results [6] , which spawned deeper investigations into printing parts with ULTEM 1010 and 9085 in this work.  ULTEM was the most radiopure starting filament analyzed (Table 3, Figure 3) with ULTEM 1010 being slightly cleaner than the 9085 variety. This suggests that the PC that is added to the formulation of ULTEM 9085 likely contains greater concentrations of Th and U or the process to create the copolymer introduces contamination, albeit at a very minute level.

**Table 3.** Average Th and U content (pg g$^{-1}$ and µBq kg$^{-1}$) of PPS filament and ULTEM 1010 and 9085 filaments and parts. All printed parts were printed using cleaned filaments.  The ULTEM 1010 dry and wet ashing results are from preliminary investigations in our previous work [6].

| Sample | Description | Digestion method | $^{232}$Th pg g$^{-1}$ | $^{232}$Th µBq kg$^{-1}$ | $^{238}$U pg g$^{-1}$ | $^{238}$U µBq kg$^{-1}$ |
|---|---|---|---|---|---|---|
| PPS | Cleaned Filament | Dry Ash | 270 ± 60 | 1100 ± 200 | 710 ± 20 | 8800 ± 300 |
| | | | | | | |
| ULTEM 1010 | Cleaned Filament | Dry Ash | 3.8 ± 1.0 | 16 ± 4 | 6.9 ± 1.5 | 85 ± 19 |
| ULTEM 1010 | Cleaned Filament | Wet Ash | 6.2 ± 0.3 | 25.4 ± 1.3 | 7.1 ± 0.3 | 87 ± 4 |
| ULTEM 1010 | Cleaned Spring Clip from Clean Filament | Dry Ash | 5.8 ± 1.2 | 25 ± 5 | 7.3 ± 1.4 | 90 ± 17 |
| | | | | | | |
| ULTEM 9085 | Cleaned Filament | Wet Ash | 9.4 ± 0.9 | 38 ± 4 | 4.5 ± 0.5 | 56 ± 6 |
| ULTEM 9085 | Cleaned Spring Clip from Clean Filament | Dry Ash | 25 ± 9 | 100 ± 40 | 7.0 ± 0.9 | 86 ± 11 |

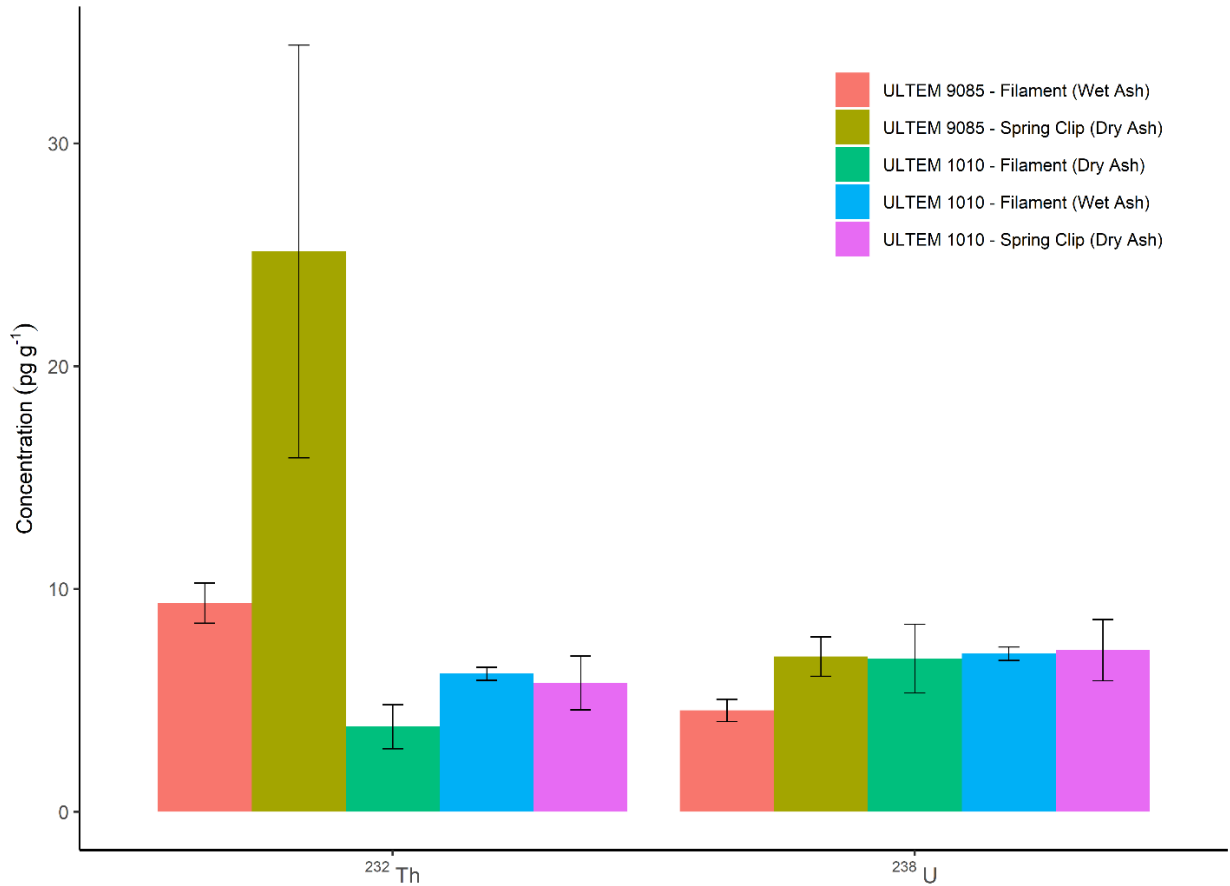

**Figure 3.** Determination of $^{238}$U and $^{232}$Th (pg g$^{-1}$) in ULTEM 9085 and 1010 filament and 3-D printed Spring Clip. The ULTEM 1010 dry and wet ashing results are from preliminary investigations in our previous work [6].

The ULTEM 1010 spring clip possesses similarly low Th and U levels to the starting filaments, showing that 3D printing can be conducted cleanly on a complex part at ultralow levels. The ULTEM 9085 had slightly elevated Th and U levels in the printed part by about a factor of 2-3X compared to the filament, albeit at still very low levels of 7 and 25 pg g$^{-1}$ for U and Th, respectively. Both types of ULTEM filaments and printed parts are significantly cleaner than the PVDF 3D printing alternatives.

*3.3 Mechanical Properties versus Radiopurity*

Using the assays reported in this paper, as well as the many assays reported in previous papers [6,7,9,11], we can chart the radiopurity of the 3D printed parts onto a graph showing mechanical properties versus radiopurity ($^{238}$U in this case), see Figure 4. This figure was adapted from our previous work [6].

Figure 4 presents tensile strength as means of assessing the mechanical properties of a range of polymers. Radiopurity is represented as the concentration of $^{238}$U on the y axis in log units, with mass concentration on the right (pg g$^{-1}$ corresponds to parts per trillion) and radioactivity concentration (microBq/kg) on the left. Radiopurity and strength are both high at the lower right-hand corner of the plot, which is the optimal location. Where a given polymer was available in multiple forms, *e.g.*, raw material as powder, or pellets, or solid, and in this case, 3D printed parts, a vertical line has been drawn to connect them. Three stars representing the work discussed here have been added for the PVDF and ULTEM 3D printed spring clip parts, showing their radiopurity in relation to other high-performance polymers. The radiopurity of 3D printed ULTEM 1010 part is within a factor of a few relative to a solid part formed using a conventional process, while the 3D printed PVDF part is nearly an order of magnitude more radiopure than a conventionally formed PVDF part. While the mechanical properties of polymers may be altered from the 3D printing process, assessing its effects on PVDF and ULTEM was beyond the scope of this study. There are a range of studies that have looked at multiple mechanical properties of ULTEM variations under different processing parameters ultimately presenting a large range in the strength of 3D printed parts that vary greatly with the processing parameters used [10,12–16]. Moreover, in some detectors, outgassing (*viz.* virtual leaks) and/or gas permeability from polymer parts are a major concern but is beyond the scope of this study for 3D printed parts.

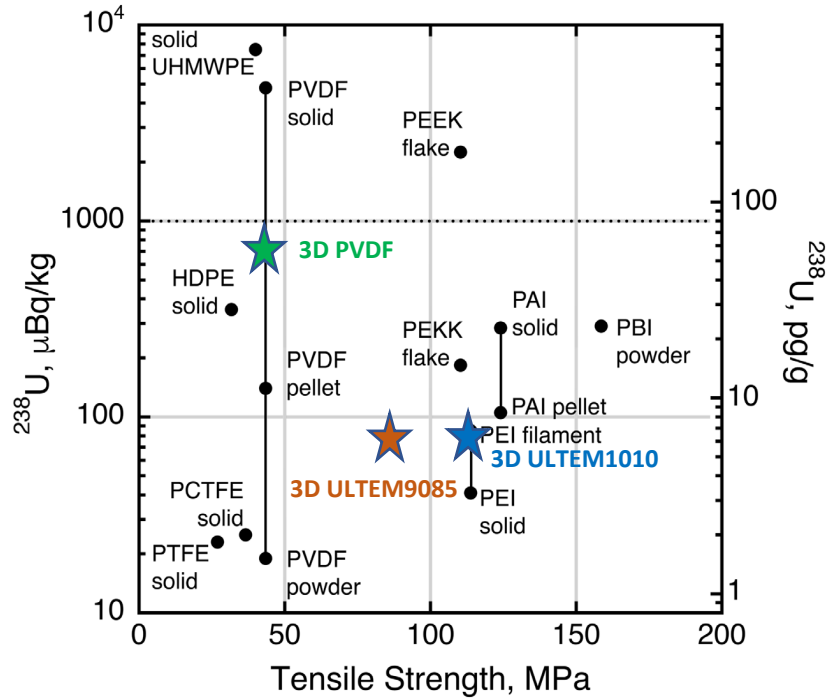

**Figure 4.** Average $^{238}$U assay values on a log scale compared to mechanical tensile strength as determined by standard ASTM test data for tensile strength (test D638). Figure was adapted from our previous work [6] with overlayed results for 3D printed spring clips parts represented by stars.

## 4. Conclusion

Additive manufacturing continues to grow in application. FDM 3D printing was explored to create ULB parts for rare event physics, allowing for a means to make complicated parts without the need for complex machining, perhaps even at the location of rare event physics experiments (*e.g.,* underground). The use of 3D printing substantially decreases the handling of intricate parts which shows promising utility in ULB physics to improve radiopurity of currently manufactured parts for a variety of detector types. Results from this study show that filaments and parts in the sub-ppt $^{232}$Th and $^{238}$U regime can be attained and produced at ultralow background levels. ULTEM 1010 and 9085 are the most radiopure in this study, at sub-100 microBq/kg levels for $^{238}$U and $^{232}$Th levels.

Future investigations should include examining the material strength of the printed polymer *after* processing to determine if the mechanical properties of the thermoplastic used remain similar post-processing. Moreover, investigations should focus on stereolithographic (SLA) methods to assess if resin bed photocuring methods can be leveraged for ULB applications.

## 5. Acknowledgments

Pacific Northwest National Laboratory (PNNL) is operated by Battelle for the United States Department of Energy (DOE) under Contract no. DE-AC05-76RL01830 . This study was supported by the DOE Office of High Energy Physics' Advanced Technology R&D subprogram.

**References**

[1] A. Ambrosi, A. Bonanni, How 3D printing can boost advances in analytical and bioanalytical chemistry, Microchimica Acta. 188 (2021). https://doi.org/10.1007/s00604-021-04901-2/Published.

[2] W. Xu, S. Jambhulkar, Y. Zhu, D. Ravichandran, M. Kakarla, B. Vernon, D.G. Lott, J.L. Cornella, O. Shefi, G. Miquelard-Garnier, Y. Yang, K. Song, 3D printing for polymer/particle-based processing: A review, Composites Part B: Engineering. 223 (2021) 109102. https://doi.org/10.1016/j.compositesb.2021.109102.

[3] K.B. Mustapha, K.M. Metwalli, A review of fused deposition modelling for 3D printing of smart polymeric materials and composites, European Polymer Journal. 156 (2021) 110591. https://doi.org/10.1016/j.eurpolymj.2021.110591.

[4] O.A. Mohamed, S.H. Masood, J.L. Bhowmik, Optimization of fused deposition modeling process parameters: a review of current research and future prospects, Advances in Manufacturing. 3 (2015) 42–53. https://doi.org/10.1007/s40436-014-0097-7.

[5] T.D. Ngo, A. Kashani, G. Imbalzano, K.T.Q. Nguyen, D. Hui, Additive manufacturing (3D printing): A review of materials, methods, applications and challenges, Composites Part B: Engineering. 143 (2018) 172–196. https://doi.org/10.1016/j.compositesb.2018.02.012.

[6] J.W. Grate, I.J. Arnquist, E.W. Hoppe, M. Bliss, K. Harouaka, M.L. di Vacri, S.A. Anguiano, Mass spectrometric analyses of high performance polymers to assess their radiopurity as ultra low background materials for rare event physics detectors, Nuclear Instruments and Methods in Physics Research, Section A: Accelerators, Spectrometers, Detectors and Associated Equipment. 985 (2021). https://doi.org/10.1016/j.nima.2020.164685.

[7] I.J. Arnquist, E.J. Hoppe, M. Bliss, J.W. Grate, Mass Spectrometric Determination of Uranium and Thorium in High Radiopurity Polymers Using Ultra Low Background Electroformed Copper Crucibles for Dry Ashing, Analytical Chemistry. 89 (2017) 3101–3107. https://doi.org/10.1021/acs.analchem.6b04854.

[8]     B.D. Laferriere, T.C. Maiti, I.J. Arnquist, E.W. Hoppe, A novel assay method for the trace determination of Th and U in copper and lead using inductively coupled plasma mass spectrometry, Nuclear Instruments and Methods in Physics Research, Section A: Accelerators, Spectrometers, Detectors and Associated Equipment. 775 (2015) 93–98. https://doi.org/10.1016/j.nima.2014.11.052.

[9]     I.J. Arnquist, E.W. Hoppe, M. Bliss, K. Harouaka, M.L. di Vacri, J.W. Grate, Mass spectrometric assay of high radiopurity solid polymer materials for parts in radiation and rare event physics detectors, Nuclear Instruments and Methods in Physics Research, Section A: Accelerators, Spectrometers, Detectors and Associated Equipment. 943 (2019). https://doi.org/10.1016/j.nima.2019.162443.

[10]    R.J. Zaldivar, D.B. Witkin, T. McLouth, D.N. Patel, K. Schmitt, J.P. Nokes, Influence of processing and orientation print effects on the mechanical and thermal behavior of 3D-Printed ULTEM ® 9085 Material, Additive Manufacturing. 13 (2017) 71–80. https://doi.org/10.1016/j.addma.2016.11.007.

[11]    D.S. Leonard, P. Grinberg, P. Weber, E. Baussan, Z. Djurcic, G. Keefer, A. Piepke, A. Pocar, J.L. Vuilleumier, J.M. Vuilleumier, D. Akimov, A. Bellerive, M. Bowcock, M. Breidenbach, A. Burenkov, R. Conley, W. Craddock, M. Danilov, R. DeVoe, M. Dixit, A. Dolgolenko, I. Ekchtout, W. Fairbank, J. Farine, P. Fierlinger, B. Flatt, G. Gratta, M. Green, C. Hall, K. Hall, D. Hallman, C. Hargrove, R. Herbst, J. Hodgson, S. Jeng, S. Kolkowitz, A. Kovalenko, D. Kovalenko, F. LePort, D. Mackay, M. Moe, M. Montero Díez, R. Neilson, A. Odian, K. O'Sullivan, L. Ounalli, C.Y. Prescott, P.C. Rowson, D. Schenker, D. Sinclair, K. Skarpaas, G. Smirnov, V. Stekhanov, V. Strickland, C. Virtue, K. Wamba, J. Wodin, Systematic study of trace radioactive impurities in candidate construction materials for EXO-200, Nuclear Instruments and Methods in Physics Research, Section A: Accelerators, Spectrometers, Detectors and Associated Equipment. 591 (2008) 490–509. https://doi.org/10.1016/j.nima.2008.03.001.

[12]    Y. Zhang, J.P. Choi, S.K. Moon, Effect of geometry on the mechanical response of additively manufactured polymer, Polymer Testing. 100 (2021) 107245. https://doi.org/10.1016/j.polymertesting.2021.107245.

[13]    C. Yang, X. Tian, D. Li, Y. Cao, F. Zhao, C. Shi, Influence of thermal processing conditions in 3D printing on the crystallinity and mechanical properties of PEEK material, Journal of Materials Processing Technology. 248 (2017) 1–7. https://doi.org/10.1016/j.jmatprotec.2017.04.027.

[14]    K.I. Byberg, A.W. Gebisa, H.G. Lemu, Mechanical properties of ULTEM 9085 material processed by fused deposition modeling, Polymer Testing. 72 (2018) 335–347. https://doi.org/10.1016/j.polymertesting.2018.10.040.

[15]    G. Taylor, S. Anandan, D. Murphy, M. Leu, K. Chandrashekhara, Fracture toughness of additively manufactured ULTEM 1010, Virtual and Physical Prototyping. 14 (2019) 277–283. https://doi.org/https://doi.org/10.1080/17452759.2018.1558494.

[16]	M.N.M. Norani, M.I.H.C. Abdullah, M.F. bin Abdollah, H. Amiruddin, F.R. Ramli, N. Tamaldin, Mechanical and tribological properties of fff 3d-printed polymers: A brief review, Jurnal Tribologi. 29 (2021) 11–30.